\documentclass[11pt,a4paper]{article}
\usepackage{jheppub}

\usepackage{amsfonts}
\usepackage{latexsym}
\usepackage{amsmath}
\usepackage{amssymb}
\usepackage{amssymb}
\usepackage{slashed}
\usepackage{upgreek}
\usepackage{mathrsfs}
\usepackage{bbm}

\usepackage{relsize}
\usepackage{graphicx}

\font\mybb=msbm10 at 11pt

\def\bb#1{\hbox{\mybb#1}}

\def\bZ {\bb{Z}}
\def\bR {\bb{R}}

\newcommand{\bea}{\begin{eqnarray}}
\newcommand{\eea}{\end{eqnarray}}

\title{ Seeking the balance: patching double and exceptional field theories}
\author{G.~ Papadopoulos,}
\affiliation{Department of Mathematics,\\ King's College London,\\
Stand, London WC2R 2LS, UK}

\emailAdd{george.papadopoulos@kcl.ac.uk}

\abstract{We investigate the patching of double and exceptional field  theories. In double field theory the patching  conditions imposed on the spacetime   after solving
the strong section condition imply that the 3-form field strength $H$ is exact. A similar conclusion can be reached for the  form field strengths of exceptional field theories after some plausive assumptions
are made on the relation between the transition functions of the additional coordinates and the patching data of the form field strengths.
We illustrate the issues that arise, and  explore several alternative  options which include the introduction  of C-folds and of the topological geometrisation condition. }

\arxivnumber{1402.2586}

\begin{document}
\maketitle
\flushbottom

\section{Introduction}

It has been known for sometime that given a spacetime $M$ and a 2-form gauge field, $F$, $dF=0$, one can introduce an additional coordinate, a charge coordinate,  and together with the Dirac quantisation condition, one can construct a $U(1)$ bundle $M_F$ over the spacetime $M$.  This bundle is a new manifold associated to the Maxwell field and the transition functions of the fibre are related to the patching conditions of the 1-form gauge potential of $F$.  This construction is a manifestation of the isomorphism of $U(1)$ bundles over $M$ with  the cohomology classes in $H^2(M,\bZ)$ and  underpins   Kaluza-Klein theory. In the latter,   the $U(1)$ gauge potential is a component
  of the  metric in higher dimensions. This is sometimes referred as the geometrisation\footnote{Later, we shall introduce a related notion referred to as ``topological
    geometrisation''.} of the $U(1)$ field. A feature of the construction is that the  charge of the $U(1)$ field is replaced by information stored in the patching conditions of $M_F$.

Similar suggestions have emerged in the context of string theory and M-theory following the early works of \cite{duff, arkady, siegel}. These include  double field theory (DFT)  of \cite{hull1, hull2, park, hohmx} applied to string theory,  the $E_{11}$  \cite{west1, west2}, and the exceptional field theory (EFT) \cite{perry3, waldram2, nicolai01, samtleben, grana} proposals,  see also reviews \cite{rev1, rev2, rev3} and references within. There are several reasons for this. One is to find a geometric realization of string and M-theory dualities and to describe duality covariant theories.
Another is to explore the idea that string and M-theory dualities emerge as symmetries of the 10- and 11-dimensional theories rather than just their toroidal compactifications following the early work of
\cite{nicolai1}.
The constructions are broadly based on a similar technology to the 2-form $U(1)$ gauge field described above but now the metric in higher dimensions is replaced by a generalized metric
which includes the form gauge potentials of string theory and M-theory, and the introduction of suitable new coordinates.

A related construction is that of generalized geometry \cite{hitchin, gualtieri}. In generalized geometry no new
coordinates are introduced in addition to those of spacetime. Instead the tangent space of the spacetime is replaced by a vector bundle $E$ which is an extension of $TM$ equipped with an appropriate bracket. Such
 an approach has been used to explore some geometric properties of supergravities associated with strings and M-theory \cite{hull3, waldram, hillmann, perry2}.

Although much work has been done to understand the  geometry that underpins  DFT and EFT several questions remain. One question is to unravel  topological and
differential structures of double and exceptional spaces, and another related question is to understand how information about  string and brane charges
is stored in their topology.  In the context of the formalism developed so far, it is not possible to answer these questions  because
most of the computations have been made using infinitesimal transformations generated by generalized Lie derivatives. However for DFT a set of finite transformations
 have been proposed \cite{hz, rev3} by integrating the infinitesimal transformations that have been known before, and have been explored as transition functions for double spaces  in
 \cite{perry1}. Infinitesimal transformations for the coordinates of exceptional spaces underlying EFT have been proposed in eg \cite{perry3, waldram2, berman, cederwall, samtleben}.

In this paper, we shall review some of the properties of patching closed form field strengths on a spacetime and illustrate the issues  involved.
 Then we shall demonstrate that the transformations of \cite{hz, rev3} that solve the strong section condition after interpreting them
 as patching conditions following \cite{perry1} imply that the  NS-NS 3-form
 field strength $H$ which arises in string theory is an exact 3-form.

In addition, we explore the double space of $T^3$ with H-flux background of \cite{rev3} from the patching conditions point of view.
We find that without additional assumptions on  the patching conditions, the construction is inconsistent as it requires that $H$ is exact. 
However, we also show that it is possible
to modify the patching conditions at triple overlaps   in such a way that consistency is restored. But such a modification is dependent
on the existence of a particular atlas on $T^3$, ie it is not general covariant,  and so it cannot be adapted to other backgrounds.

We shall further argue that a similar conclusion can be reached in the context of EFT provided that the patching conditions
of closed form field strengths are related in a linear way to the transition functions of the additional coordinates of the
exceptional spaces and the combinatorial law follows the usual rules of tensor calculus. The latter point will be illustrated in the
$U(1)$ paradigm reviewed in the next section. Concerns about the consistency of patching
EFT have also been raised in \cite{nicolai2} using the Cartan theorem \cite{cartan}.

 We shall explore several alternative ways to reconcile the transition functions of the double and exceptional spaces with the patching conditions
 of the closed form field strengths on the spacetime. We shall see that there are examples described in appendix A, where this can be done at a cost. In particular, these constructions do not exhibit
  some key properties  of the $U(1)$ paradigm. Nevertheless,
they illustrate some of the issues involved and provide a local model of a consistent construction.

In the conclusions, we propose a general scheme with a minimum number
 of requirements that should be followed in order to construct double and exceptional spaces which exhibit the key properties of the $U(1)$ paradigm and allow for a realization
 of duality groups. To distinguish them
 from previous constructions, we shall call these new spaces ``C-folds'', ie manifolds with charge coordinates.
  One of the key requirements that we propose is  the ``topological geometrisation condition''. This states that the pull back on the C-fold of the closed form field strengths on a spacetime
 must be exact. One of the consequences of this condition is that the fibres  of the C-fold must have a non-trivial topology. We propose several constructions for C-folds based on K-theory
 and homotopy theory.

This paper has been organized as follows. In section 2, we describe the $U(1)$ field paradigm. In section 3, we show that for DFT the patching conditions allow
for only exact form field strengths. I section 4, we revisit the construction of the double space of $T^3$ with $H$-flux background.
 In section 5, we show under certain assumptions a similar statement for the EFT. In section 6, we give our conclusions and explore some future directions.
In appendix A, we give an example of an alternative construction for  double and exceptional geometries consistent with closed
form field strengths.

\section{A review of  $U(1)$ field paradigm}\label{2formpara}

Before we proceed to investigate how to extend a spacetime with additional coordinates associated to a general  closed k-form, it is instructive to
revisit  the construction for $U(1)$ fields  which is well known.

To begin, suppose $M$ is a n-dimensional manifold equipped with a good cover $\{U_\alpha\}_{\alpha\in I}$, see eg \cite{bott},  and a closed 2-form $\omega^2$. On each open set
$U_\alpha$, the Poincar\'e theorem implies that there are 1-form potentials $C^1_\alpha$ such that
\bea
\omega^2_\alpha= dC^1_\alpha~.
\eea
Using repeatedly the Poincar\'e lemma at double and triple overlaps $U_\alpha\cap U_\beta$ and $U_\alpha\cap U_\beta\cap U_\gamma$, one has that
\bea
C_\alpha^1-C_\beta^1=d a^0_{\alpha\beta}~,~~~a^0_{\alpha\beta}+a^0_{\gamma\alpha}+a^0_{\beta\gamma}=n_{\alpha\beta\gamma}~,
\label{over2}
\eea
where $n_{\alpha\beta\gamma}$ are constants. Now if $\omega^2$ represents a class in $H^2(M, \bZ)$, then $n_{\alpha\beta\gamma}\in 2\pi \bZ$ are integers.

The geometrisation of the $U(1)$ field proceeds as follows. Starting from the good cover  $\{U_\alpha\}_{\alpha\in I}$ of $M$, one introduces a new space
 with charts $U_\alpha\times \bR$, and transition functions
\bea
x_\alpha=f_{\alpha\beta}(x_\beta)~,~~~\theta_\alpha=\theta_\beta+ a^0_{\alpha\beta}~,
\label{tras2}
\eea
where $x_\alpha$ and $f_{\alpha\beta}$ are the coordinates and transition functions of $M$ and $\theta_\alpha$ is an additional coordinate.
At first sight it appears that the second transition function  is consistent with (\ref{over2}) at triple overlaps iff $n_{\alpha\beta\gamma}=0$. As we shall prove later,  this implies that $\omega^2$ is exact. However,
 for any closed $\omega^2$  which represents a class in $H^2(M, \bZ)$, the transition functions are consistent at triple overlaps by  taking  the addition in the second transition function  in (\ref{tras2}) to be
over mod $2\pi\bZ$, ie the new transition functions are
\bea
x_\alpha=f_{\alpha\beta}(x_\beta)~,~~~(\theta_\alpha-\theta_\beta- a^0_{\alpha\beta})=0~\mathrm{mod}\, 2\pi \bZ~.
\label{tras21}
\eea
In such a case no further condition arises as the sum over $a^0$'s in (\ref{over2}) at triple overlaps vanishes mod $2\pi\bZ$.

The effect of this construction  is to define  a circle bundle  $M_{\omega^2}$ over $M$ with transition functions $\phi_{\alpha\beta}=\exp (i a^0_{\alpha\beta})$. Some of the features and  consequences
of this construction  are as follows.

\begin{itemize}

\item{(i)} An integral part of the construction is the modification of the combinatorial law that it is used to describe the transition functions of the  additional coordinate.  This is related to
the requirement that  $\omega^2$ represents a class in $H^2(M, \bZ)$.

\item{(ii)} The topological structure of $M_{\omega^2}$ is completely determined by the class $[\omega^2]$ of $\omega^2$ in $H^2(M, \bZ)$, ie $M_{\omega^2}=M_{[\omega^2]}$,
and vice versa, as there is a 1-1 correspondence between circle bundles and elements of $H^2(M, \bZ)$.

\item{(iii)} The tangent bundle $TM_{\omega^2}$ of $M_{\omega^2}$ is an extension of $TM$ with respect to a trivial real line bundle $L$, ie
\bea
0\rightarrow L \rightarrow TM_{\omega^2}\rightarrow \pi^* TM\rightarrow 0
\eea
where $\pi : M_{\omega^2}\rightarrow M$ is a projection.

\item{(iv)} Another feature of the construction is that $\pi^* \omega^2$ is an exact form on $M_{\omega^2}$ as
\bea
\pi^* \omega^2=- d (d\theta- \pi^* C^1)
\eea
and $(d\theta- \pi^* C^1)$ is a globally defined 1-form on $M_{\omega^2}$.

\end{itemize}

The last property can be seen as the ``topological geometrisation'' of  $\omega^2$. On $M$, $\omega^2$ has charges which are given by the periods
\bea
n_i=\int_{B_i} \omega^2
\eea
where $(B_i)$ is a basis in $H_2(M, \bZ)$. Since  $\pi^*\omega^2$ is exact on $M_{\omega^2}$, all the periods of $\pi^*\omega^2$ on $M_{\omega^2}$ vanish. Instead  all the information carried by the periods $n_i$ has been replaced
by the topology of $M_{\omega^2}$.

The above construction can be generalized to include more than one  $U(1)$ field strengths leading to  toric fibrations over $M$. In addition
this can be generalized to non-abelian gauge fields which in turn leads to principal bundles with fibre the gauge group. In both cases
all the properties mentioned above, with some minor modifications, apply to the more general set up.

\section{Patching 3-forms and double coordinates}

\subsection{Transition functions of closed 3-forms}
\label{patch32}

Before, we proceed to the patching of k-forms and the introduction of new coordinates, it is instructive to explain the patching of closed 3-forms $\omega^3$.
 For this let $M$ be a n-dimensional manifold\footnote{In what follows, the signature of $M$ is not essential and the results apply to both Riemannian and Lorentzian manifolds.} with a good cover  $\{U_\alpha\}_{\alpha\in I}$ and a partition of unity $\{\rho_\alpha\}_{\alpha\in I}$
subordinate to $\{U_\alpha\}_{\alpha\in I}$. This means that $M$ admits functions $\rho_\alpha\geq 0$ with support in $U_\alpha$ such that at every point $x\in M$
\bea
\sum_\alpha \rho_\alpha=1~,
\label{part1}
\eea
where the sum is taken over a finite collection. For a discussion on  partitions of unity and the different kinds that exist see \cite{bott}. Here we shall use those partitions of unity
that have the same index set $I$ as that of the good cover and so they do not have necessarily compact support.

To continue, take $\omega^3$ to be a closed 3-form on  $M$. Using the Poincar\'e lemma at each open set $U_\alpha$, we can write
$\omega^3$ in terms of a gauge potential $C$ as
\bea
\omega^3_\alpha=d C^2_\alpha~.
\eea
Using repeatedly the Poincar\'e lemma  at the double,  $U_\alpha\cap U_\beta$, triple, $U_\alpha\cap U_\beta\cap U_\gamma$, and quadruple,  $U_\alpha\cap U_\beta\cap U_\gamma\cap U_\delta$, overlaps,
we have
\bea
C^2_\alpha= C^2_\beta+d a^1_{\alpha\beta}~,~~~a^1_{\alpha\beta}+ a^1_{\beta\gamma}+a^1_{\gamma\alpha}=d a^0_{\alpha\beta\gamma}~,~~~ a^0_{\alpha\beta\gamma}- a^0_{\delta\alpha\beta}
+a^0_{\gamma\delta\alpha}-a^0_{\beta\gamma\delta}=n_{\alpha\beta\gamma\delta}~,
\label{tran3}
\eea
where $n_{\alpha\beta\gamma\delta}$ are constants.  Note that the patching data $a^1_{\alpha\beta}, a^0_{\alpha\beta\gamma}, n_{\alpha\beta\gamma\delta}$ are skew-symmetric in the interchange of any two
of the open set labels, ie $a^1_{\beta\gamma}=-a^1_{\gamma\beta}$ and similarly for the rest. Moreover $n_{\alpha\beta\gamma\delta}$  are restricted to be multiples of integers if $\omega^k$ represents a class in $H^3(M, \bZ)$. In particular, the patching
condition of 2-form gauge potential in local coordinates reads
\bea
(C^2_\alpha)_{ij}&=& {\partial x_\beta^k\over \partial x_\alpha^i} {\partial x_\beta^l\over \partial x_\alpha^j} (C_\beta^2)_{kl} +  {\partial\over\partial x_\alpha^i} a^1_{\alpha\beta, j}-
 {\partial\over\partial x_\alpha^j} a^1_{\alpha\beta, i}
 \cr
 &=& {\partial x_\beta^k\over \partial x_\alpha^i} {\partial x_\beta^l\over \partial x_\alpha^j} \Big((C^2_\beta)_{kl}+
 {\partial\over\partial x_\beta^k} a^1_{\alpha\beta, l}-
 {\partial\over\partial x_\beta^l} a^1_{\alpha\beta, k}\Big)~.
\eea

Observe that the choice of gauge potentials is not unique. In particular, they are defined up to  the gauge transformation
\bea
C^2_\alpha\rightarrow C^2_\alpha+ d \chi^1_\alpha~.
\eea
Similarly, the patching data  $a^1_{\beta\gamma}$  of $\omega^3$ at double overlaps are not uniquely defined either. In particular, they are defined up to  a gauge transformation
\bea
a^1_{\alpha\beta}\rightarrow a^1_{\alpha\beta}-\chi^1_\alpha+\chi^2_\beta+ d \psi^0_{\alpha\beta}~,
\label{amb3}
\eea
for a 0-form $\psi^0_{\alpha\beta}$ defined on double overlaps. This is the only ambiguity that one
has in determining the patching conditions on double overlaps of $\omega^3$. Anything else is inconsistent with
the identification of $\omega^3$ as a closed 3-form on $M$.

\subsection{Transition functions and exact 3-forms}
\label{exact3}

Before we proceed to compare the above patching conditions of closed 3-forms with those that arise in DFT, we shall prove a technical
lemma which arises in the context of $\check{\mathrm C}$ech-de Rham theory. In particular,
 if there is a choice of patching data, up to (\ref{amb3}) gauge transformations, such that $a^1_{\alpha\beta}$  on triple overlaps $U_\alpha\cap U_\beta\cap U_\gamma$
satisfies the cocycle\footnote{This is in the sense of $\check{\mathrm C}$ech cohomology, see \cite{bott}.} condition
\bea
a^1_{\alpha\beta}+ a^1_{\beta\gamma}+a^1_{\gamma\alpha}=0~,
\label{cocycle1}
\eea
then $\omega^3$ is exact.

For this suffices to show that $\omega^3=d \tilde C^2$, where $\tilde C^2$ is a 2-form on $M$, ie $\tilde C_\alpha=\tilde C_\beta$.
Indeed define
\bea
\tilde C^2_\alpha=C^2_\alpha+\sum_\gamma d (\rho_\gamma  \, a^1_{\gamma\alpha})~.
\eea

Observe that $d\tilde C_\alpha= dC_\alpha= \omega^3$ and
\bea
\tilde C_\alpha-\tilde C_\beta&=&C^2_\alpha-C^2_\beta+\sum_\gamma d (\rho_\gamma\,  a^1_{\gamma\alpha})-\sum_\gamma d (\rho_\gamma\,   a^1_{\gamma\beta})
\cr
&=&da^1_{\alpha\beta}-\sum_\gamma d(\rho_\gamma\,  a^1_{\alpha\beta})=da^1_{\alpha\beta}-da^1_{\alpha\beta}=0~,
\eea
where to establish the last two equalities we have used (\ref{cocycle1}) and (\ref{part1}), respectively.
Thus if (\ref{cocycle1}) holds,  $\omega^3$ is exact and represents the trivial class in $H^3(M, \bR)$.

\subsection{Patching DFT}
\label{pdft}

After integrating the infinitesimal transformations generated by generalized Lie derivatives, generalized finite tensor transformations have  been proposed for DFT\footnote{For some mathematical aspects of DFT see eg \cite{vaisman} and also \cite{kikuchi} for explicit backgrounds.} in  \cite{hz, rev3} and further explored
 as transition functions in \cite{perry1}. According to these generalized tensor transformations,
1-forms are transformed as
\bea
\omega'_N= F_N{}^M \omega_M~,
\eea
where
\bea
F_N{}^M= {1\over2} \big({\partial X^P\over\partial X'^N} {\partial X'_P\over\partial X_M}+ {\partial X'_N\over\partial X_P } {\partial X^M\over\partial X'^P}\big)~,
\eea
and where $X^M$ are the  coordinates of the double space  and indices are raised and lowered with the split metric $\eta_{MN}$ and $X'^M=X'^M(X^N)$.
All the fields and coordinate transformations satisfy the strong section condition.

Viewing the above transformations as patching conditions and putting them into the language of the previous section, we have
\bea
X^M_\alpha=X^M_{\alpha\beta}(X_\beta^N)~,~~~(\omega_\alpha)_N= (F_{\alpha\beta})_N{}^M (\omega_\beta)_M~,
\eea
where
\bea
(F_{\alpha\beta})_N{}^M={1\over2} \big({\partial X_\beta^P\over\partial X_\alpha^N} {\partial X_{\alpha P}\over\partial X_{\beta M}}+ {\partial X_{\alpha N}\over\partial X_{\beta P} } {\partial X_\beta^M\over\partial X_\alpha^P}\big)~.
\label{genpatch}
\eea

An extensive investigation of the transformations induced on the fields after solving the strong section condition\footnote{If the strong section condition is relaxed, then the generalized transformations
 on the forms do not commute with the exterior derivative.  So closed forms do not transform
 to closed forms.}has been made in \cite{hz, rev3}.
Interpreting these transformations as patching conditions,
one can show that  if one  considers either
\bea
x^i_\alpha= x^i_{\alpha\beta}(x_\beta^j),~~~~ y_{\alpha i}=y_{\beta i}
\eea
or shift transformations
\bea
x^i_\alpha=x^i_\beta, ~~~~y_{\alpha i}=y_{\beta i}-\zeta_{\alpha\beta i}
\label{sdcor}
\eea
 of the double coordinates,
then the  2-form gauge potential $b$ of the NS-NS 3-form field strength transforms either as a 2-form, $b_\alpha=b_\beta$ or as $b_\alpha=b_\beta+ d\zeta_{\alpha\beta}$, where
in the latter case the $x$ transformation is the identity

Clearly, none of these two patching conditions are satisfactory. If $b$ transforms as a 2-form, then one has to restrict $H$ to be exact. The shift transformation is also not sufficient
as we know that string theory admits solutions which require that $M$ patches with non-trivial transition functions.

The authors of \cite{rev3} have also investigated the combined transformation which includes diffeomorphisms of the spacetime $M$ combined with shift transformations
of the $y$ coordinates which as a patching condition reads
\bea
x^i_\alpha= x^i_{\alpha\beta}(x_\beta^j)~,~~~y_{\alpha i}=y_{\beta i}-\zeta_{\alpha\beta i}~. \label{patch33}
\eea
In turn these give rise to the patching condition
\bea
(b_\alpha)_{ij}&=& {\partial x_\beta^k\over \partial x_\alpha^i}\, {\partial x_\beta^l\over \partial x_\alpha^j} \, \Big((b_\beta)_{kl}+{1\over2}\big(
 {\partial\over\partial x_\beta^k} \zeta_{\alpha\beta, l}-
 {\partial\over\partial x_\beta^l} \zeta_{\alpha\beta, k}\big)\Big)
 \cr
 &&~~~+{1\over2} \Big( {\partial x_\beta^k\over \partial x_\alpha^i}\, {\partial\over\partial x_\beta^k} \zeta_{\alpha\beta, j} -
   {\partial x_\beta^k\over \partial x_\alpha^j} {\partial\over\partial x_\beta^k} \zeta_{\alpha\beta, i}\Big)~,
   \label{bpatch}
\eea
for the 2-form gauge potential $b$.

It is clear from (\ref{patch33}) that consistency  on triple overlaps on the spacetime requires that
\bea
\zeta_{\alpha\beta }+\zeta_{\beta\gamma }+\zeta_{\gamma\alpha }=0~.
\eea
If this is the case, a consequence of the lemma proven in section \ref{exact3} implies that there are 1-forms $\{u_\alpha\}$ defined on the open sets $\{U_\alpha\}$ such that
\bea
\zeta_{\alpha\beta}= -u_\alpha+u_\beta~,
\eea
where
\bea
u_\alpha=\sum \rho_\gamma \zeta_{\gamma\alpha}~.
\eea

Thus (\ref{bpatch}) can be rewritten as
\bea
\big((b_\alpha)_{ij} &+&{1\over2} (du_\alpha)_{ij} \big)\, dx_\alpha^i\wedge dx_\alpha^j=\big((b_\beta)_{ij} +{1\over2} (du_\beta)_{ij} \big)\, dx_\beta^i\wedge dx_\beta^j
\cr
&-&{1\over2}{\partial\over \partial x^i_\beta}\big(u_{\alpha k} {\partial x^k_\alpha\over \partial x_\beta^j}\big) dx_\beta^i\wedge dx_\beta^j+{1\over2}{\partial\over \partial x^i_\beta}\big( u_{\beta k} {\partial x^k_\alpha\over \partial x_\beta^j}\big) dx_\beta^i\wedge dx_\beta^j~,
\eea
or equivalently,
\bea
b_\alpha-b_\beta= -{1\over2} du_\alpha+ {1\over2} du_\beta-{1\over2} d\tilde u_\alpha+{1\over2} d\tilde u_\beta~.
\eea
 As a result
 \bea
\tilde b_\alpha\equiv b_\alpha+{1\over2} du_\alpha+{1\over2} d\tilde u_\alpha=b_\beta+{1\over2} du_\beta+{1\over2} d\tilde u_\beta\equiv\tilde b_\beta~.
 \eea
 So  up to a gauge transformation, $b$ can be made globally defined on $M$ and $H=db=d\tilde b$. Thus   $H$ is exact.

One therefore concludes that the patching conditions induced on the spacetime from the generalized coordinate transformations of DFT after solving the strong section condition
 imply that $H$ is an exact form.

This conclusion cannot be  satisfactory  as we know string theory
has as solutions spacetimes that admit non-trivial patching and a closed but not exact form $H$. The restriction that $H$ is exact is a direct consequence
of the introduction of the new coordinates $y$, their transition functions and their relation to the patching conditions of the 2-form gauge potentials as this is implied from the generalized coordinate transformations.

It is worth pointing out that $H$ is not restricted to be exact in the context of generalized geometry\footnote{Generalized geometry does not require that $H$ represents
a class in $H^3(M, \bZ)$ and so it does not capture the analogue of the Dirac quantisation condition in string theory and M-theory. It also
does not obey the topological geometrisation condition
unlike the $U(1)$ field paradigm.}.  In generalized geometry, there are no additional coordinates that
have to be patched. The analogous  consistency condition which arises on triple overlaps  reads
\bea
d(a^1_{\alpha\beta}+a^1_{\beta\gamma}+a^1_{\gamma\alpha})=0~,
\eea
and it does not impose additional restrictions on the transition functions of $H$.
The above condition is always satisfied as it can be seen from the analysis of the  section \ref{patch32}.

\subsection{Seeking a consistent patching}

It is not apparent how to reconcile the transition functions of the double space with those of the spacetime, and the patching conditions of the 3-form field strength $H$ without imposing
additional conditions on the fluxes. Any choice of transition functions for the double space of the type
\bea
x^i_\alpha= x^i_{\alpha\beta}(x_\beta^j)~,~~~y^1_{\alpha i}=y^1_{\beta i}-\zeta^1_{\alpha\beta i}
\eea
where the patching conditions $a^1_{\alpha\beta}$ of $H$ are linear combination of  $\zeta_{\alpha\beta}^1$'s  and the combinatorial rules denoted are as those of tensor calculus, will lead to the conclusion that $H$
  is exact. Another indication that such a direct approach may not be fruitful is that there is nowhere use of the quantisation condition $[H]\in H^3(M, \bZ)$ which
  has a central role in the exploration of the $U(1)$ paradigm.

In appendix A, a modification of the patching conditions of the $y$ coordinates was proposed and it was shown that
there are no additional restrictions at triple overlaps. If this modification is considered, the double space patches consistently
and it has some attractive features like its tangent bundle is an extension of the tangent bundle of the spacetime.
However, the topological geometrisation of $H$ fails, ie when $H$ is pulled back on the double space
is not exact which is in conflict with the $U(1)$ field paradigm. Perhaps this is not surprising as to be able to topologically geometrize
$H$, the additional coordinates have to exhibit non-trivial topology related to the class of $H$ in $H^3(M, \bZ)$. In the conclusions,
we propose a general framework where all these questions may be addressed.

\section{Revisiting examples}

\subsection{Reviewing the double space of 3-torus with constant $H$ flux}

In this section, we review the construction of the double space of  $T^3$ with constant $H$ flux.
This example is the most relevant one of the three examples presented in \cite{rev3} to explore the patching of the double spaces.  This
is because the consistency of the construction can be checked at the level of the transition functions for the coordinates of the double space.
In the other two examples instead, the focus is on the consistency of the patching of generalized tensors.

After some  relabeling of the coordinates and in geometric units, the model can be described as follows. We equip $T^3$ with three angular ``coordinates''\footnote{Note these are not coordinates
 in the sense of manifold theory. Instead they should be thought as labels that denote the points on the circle.}$(\theta^1, \theta^2, \theta^3)$
such that $0\leq \theta^1, \theta^2, \theta^3<2\pi$ and take as an $H$ flux $H=N\, d\theta^1\wedge d\theta^2\wedge d\theta^3$. The quantization
of flux
\bea
{1\over (2\pi)^2} \int H\in \bZ
\eea
 requires that $2\pi N\in \bZ$.

The construction of double space in \cite{rev3} proceeds as follows. First the Poincar\'e lemma is used to locally solve  for the gauge
potential $b$ of $H$ as
\bea
b=N\, \theta^3\, d\theta^1\wedge d\theta^2~.
\eea
Then it is noticed that
\bea
b(2\pi)-b(0)= 2\pi N d\theta^1\wedge d\theta^2~.
\eea
This apparent lack of periodicity for $b$ is compensated by a gauge transformation $b'=b-d\xi$, where
\bea
\xi=2\pi N \theta^1 d\theta^2~,
\eea
and so
\bea
b'(2\pi)=b(2\pi)-2\pi N d\theta^1\wedge d\theta^2=b(0)~.
\eea
This gauge transformation is then used to define the coordinate transformations of the DFT background as
\bea
&&{\theta'}^{1}=\theta^1~,~~~{\theta'}^2=\theta^2~,~~~{\theta'}^3=\theta^3~,
\cr
&&{\psi'}^1=\psi^1~,~~~{\psi'}^2=\psi^2-2\pi N \theta^1~,~~~{\psi'}^3=\psi^3~,
\label{dualcon}
\eea
according to (\ref{sdcor}), where $(\psi^1, \psi^2, \psi^3)$ are the dual coordinates which are taken as the coordinates of a dual torus $\tilde T^3$.

Below we shall demonstrate that the construction of the double space  relies for consistency
on the existence of a particular atlas on $T^3$.  As a result, it has limited applicability. In the process of proving this, several silent
features of the above construction will become apparent.

\subsection{Patching the double space of 3-torus with constant $H$}

\subsubsection{Patching conditions}

To describe the construction of double spaces from a patching point of view, we shall first describe an atlas on the circle $S^1$ which in turn
will induce an atlas on the 3-torus $T^3$.  For the former, we  cover $S^1$ with two patches $\{(U_1, \varphi_1), (U_2, \varphi_2)\}$ such that
\bea
&&\varphi_1: U_1\subset S^1\rightarrow (-{\pi\over4}, {5\pi\over4})~,~~~~\varphi_1((s,t))=x_1~,
\cr
&&\varphi_2: U_2\subset S^1\rightarrow (-{5\pi\over4}, {\pi\over4})~,~~~~\varphi_2((s,t))=x_2~,
\label{atlas1}
\eea
where $U_1$ is the open set which includes the north pole and $U_2$ is the open set which includes the south pole, and we have solved the algebraic equation $s^2+t^2=1$ of $S^1$ as $s=\cos x_1, t=\sin x_1$ and $s=\cos x_2, t=\sin x_2$.

The intersection $\varphi_1(U_1\cap U_2)=(-{\pi\over4}, {\pi\over4})\cup ({3\pi\over4}, {5\pi\over4})$ or equivalently $\varphi_2(U_1\cap U_2)=(-{\pi\over4}, {\pi\over4})\cup (-{5\pi\over4}, -{3\pi\over4})$. Therefore the transition functions are $x_2=\varphi_2\circ \varphi_1^{-1}(x_1)$ which give
\bea
x_2&=&x_1~,~~~~\mathrm{on}~~~(-{\pi\over4}, {\pi\over4})~,
\cr
x_2&=&x_1-2\pi~,~~~\mathrm{on}~~~({3\pi\over4}, {5\pi\over4})~.
\label{atlas2}
\eea
It is convenient to write the above transition function as
\bea
x_2=x_1+2 n_x \pi~,
\eea
with the understanding that $n_x=0$ on $(-{\pi\over4}, {\pi\over4})$ and $n_x=-1$ on $({3\pi\over4}, {5\pi\over4})$. Of course the transition
function $\varphi_1\circ \varphi_2^{-1}$ is the inverse of $\varphi_2\circ \varphi_1^{-1}$, ie $x_1=x_2-2 n_x \pi$. Observe that
\bea
dx_1=dx_2~,
\eea
and so $dx$ is a globally defined closed but not exact 1-form on $S^1$ as expected. We refer to $x$ as the angular manifold coordinates of $S^1$ to distinguish them from the $\theta$s
of the previous section.

Using the above atlas on $S^1$, one can induce a manifold structure on $T^3$ as follows. First, we introduce two patches as above
for each one of the three circles, ie $\{(U_i, \varphi_i)\vert i=1,2\}$, $\{(V_j, \lambda_j)\vert j=1,2\}$ and $\{(W_k, \mu_k)\vert k=1,2\}$, and then take their products. The atlas on $T^3$ is $\{ (U_i\times V_j\times W_k, \varphi_i\times \lambda_j\times \mu_k\})\vert i,j,k=1,2 \}$, ie
$T^3$ is covered by 8 patches.  It is convenient to define $U_{ijk}= U_i\times V_j\times W_k$ and $\varphi_{ijk}=\varphi_i\times \lambda_j\times \mu_k$.

In this atlas on $T^3$, the $H$ flux can now be written as $H=N dx\wedge dy\wedge dz$, where $x, y$ and $z$ are  the
angular manifold coordinates for the three circles.  To continue we use the Poicar\'e lemma for each of the open sets $U_{ijk}$  to identify the
gauge potential of $H$ as
\bea
b_{ijk}={N\over3} (x_i\, dy\wedge dz-y_j\, dx\wedge dz+ z_k\, dx\wedge dy) \, .
\eea
Note that  the patch Greek labels $\alpha$ of section \ref{patch32}  have been replaced with the multi-labels $ijk$.
On the double overlaps  $U_{i_1j_1k_1}\cap U_{i_2j_2k_2}$, we have that
\bea
b_{i_1j_1k_1}&=&b_{i_2j_2k_2}
+{N\over 3} 2\pi [n_x (i_1-i_2) dy\wedge dz
\cr&& ~~- n_y (j_1-j_2) dx\wedge dz+ n_z (k_1-k_2) dx\wedge dy]~.
\eea
As a result, the 1-form transition functions $a^1_{i_1j_1k_1,i_2j_2k_2}$ of section \ref{patch32} can be chosen as
\bea
a^1_{i_1j_1k_1,i_2j_2k_2}&=&{N\over12} 2\pi \{[n_x (i_1-i_2) (y_{j_1}+y_{j_2}) -n_y (j_1-j_2) (x_{i_1}+x_{i_2})] dz\cr
&&+[-n_x (i_1-i_2) (z_{k_1}+z_{k_2})+ n_z (k_1-k_2) (x_{i_1}+x_{i_2})] dy
\cr &&
[n_y (j_1-j_2) (z_{k_1}+z_{k_2})- n_z (k_1-k_2) (y_{j_1}+y_{j_2})] dx\}~.
\eea
This expression respects the symmetries of the transition functions that arise from the exchange of the patch labels.

To continue, we compute $d a^0$ at triple overlaps to find that
\bea
d a^0_{i_1j_1k_1, i_2j_2k_2, i_3j_3k_3}&=&{N\over12} (2\pi)^2 \{n_x n_y [i_1(j_2-j_3)-j_1(i_2-i_3)-i_2(j_1-j_3)
\cr &&
+j_2(i_1-i_3)+ i_3 (j_1-j_2)-j_3(i_1-i_2)] dz
\cr
&&+ n_x n_z [ -i_1(k_2-k_3)+ i_2 (k_1-k_3)+ k_1 (i_2-i_3)
\cr
&&-k_2 (i_1-i_3)- i_3(k_1-k_2)+ k_3(i_1-i_2)] dy
\cr
&&
+n_y n_z [j_1 (k_2-k_3)-j_2 (k_1-k_3)- k_1 (j_2-j_3)
\cr
&&+ k_2 (j_1-j_3)- k_3 (j_1-j_2)+ j_3 (k_1-k_2)] dx\}~,
\label{t3a0}
\eea
and $n$ at quadruple intersections to find that
\bea
&&n_{i_1j_1k_1, i_2j_2k_2, i_3j_3k_3, i_4j_4k_4}={N\over24} (2\pi)^3 n_x n_y n_z \{ [(i_1j_2-j_1i_2) (k_3-k_4)+ (j_1 i_3-j_3 i_1) (k_2-k_4)
\cr
&&~~~+
(i_2 j_3-j_2 i_3) (k_1-k_4)+ (j_2 i_4-i_2 j_4) (k_1-k_3)+(j_4 i_1-j_1 i_4) (k_2-k_3)+ (i_3 j_4-j_3 i_4) (k_1-k_2)]
\cr&&~~~
-
[(i_1 k_2-k_1i_2) (j_3-j_4)+ (k_1 i_3-k_3 i_1) (j_2-j_4)+(i_2 k_3-k_2 i_3) (j_1-j_4)+ (k_2 i_4-i_2 k_4) (j_1-j_3)
\cr &&~~~
+(k_4 i_1-k_1 i_4) (j_2-j_3)
+ (i_3 k_4-k_3 i_4) (j_1-j_2)]
+ [ (j_1 k_2-k_1 j_2) (i_3-i_4)+ (k_1 j_3-k_3 j_1) (i_2-i_4)
\cr&&~~~
+ (j_2 k_3-k_2 j_3) (i_1-i_4)
+ (k_2 j_4-k_4 j_2) (i_1-i_3)
+(k_4 j_1-k_1 j_4) (i_2-i_3)
\cr&&~~~
+ (j_3 k_4-k_3 j_4) (i_1-i_2)]\}~.
\eea
 Note that $d a^0$ has constant components. This is an atlas dependent statement.  $da^0$ will not have constant
components in another atlas. Of course $n$ is constant as expected independent of the choice of atlas.

Denoting the double coordinates with $X,Y, Z$, one can according to (\ref{sdcor}) impose the patching condition
\bea
X_{i_1j_1k_1} dx+Y_{i_1j_1k_1} dy+ Z_{i_1j_1k_1} dz&=&X_{i_2j_2k_2} dx+Y_{i_2j_2k_2} dy
\cr
&&~~~+ Z_{i_2j_2k_2} dz- a^1_{i_1j_1k_1,i_2j_2k_2}~.
\label{npatcon}
\eea
Without further assumptions on $X,Y,Z$, a check of the consistency of the above patching condition at triple overlaps will lead to the requirement that
\bea
da^0=0~.
\eea
This is clearly  a contradiction as $da^0\not=0$ which  can be seen from  (\ref{t3a0}). This is in agreement
with the proof presented in sections \ref{patch32}, \ref{exact3} and  \ref{pdft}.

However for this particular example, one can also impose as patching condition
\bea
&&\Big(X_{i_1j_1k_1} dx+Y_{i_1j_1k_1} dy+ Z_{i_1j_1k_1} dz-X_{i_2j_2k_2} dx-Y_{i_2j_2k_2} dy
\cr
&&~~~~
- Z_{i_2j_2k_2} dz+ a^1_{i_1j_1k_1,i_2j_2k_2}\Big)=0~~ \mathrm{mod}\,~~\bZ (\ell, \ell,\ell) ~,
\label{3tras}
\eea
instead of that in (\ref{npatcon}).
Choosing $\ell={N\over3} (2\pi)^2 $, the inconsistency at the triple overlaps is removed and  the patching  becomes consistent.

Note however that this modification of the transition functions at triple overlaps goes beyond the suggestion of \cite{perry1}
that the transition functions are just the local diffeomorphisms of the double space. To put it in another way, the periodic identification
of the dual coordinates does not follow from the patching conditions of the double space alone as constructed from the patching conditions
of $T^3$ and the 2-form gauge potential. Instead it has to be imposed by hand.

It remains to investigate whether a modification of the patching condition as in (\ref{3tras}) is always possible.
This will be explored below.

\subsubsection{Consistency of the patching}

The modification of the patching condition as in (\ref{3tras}) comes at a cost.
 First,
the construction is atlas dependent. If one uses another atlas on $T^3$, the components of $da^0$ are functions of the coordinates
instead of constants.  As a result, the modification of the patching condition as in (\ref{3tras}) is not valid.

Nevertheless, one could argue that since there is an atlas on $T^3$ as in (\ref{atlas1}) and (\ref{atlas2}) that the patching condition (\ref{3tras})
is valid, one can perform a diffeomorphism on $T^3$ which can take $T^3$ with respect to any atlas to $T^3$ with the (\ref{atlas1}) and (\ref{atlas2}) atlas.
Such a diffeomorphism  $x'=x'(x)$ will transform the dual coordinates schematically as
\bea
X'={\partial x\over \partial x'} X~,~~~Y'={\partial x \over \partial x'} Y~,~~~Z'={\partial x \over \partial x'} Z~.
\eea
However, such a transformation is not allowed within the DFT as does not solve the strong section condition.

Furthermore it is worth contrasting the patching conditions (\ref{3tras}) with those of (\ref{tras21})  in section \ref{2formpara} for  the 2-form paradigm.
 The modification of the combinatorial law in  (\ref{tras21})
and the subsequent consistency of patching at triple overlaps
are atlas independent.  As a result, the mechanism in section \ref{2formpara} can apply to any background and any closed 2-form provided that represents an
integral class. This is not the case for the patching in (\ref{3tras}) as we have seen.

The above comparison also explains the difficulty in constructing the double spaces of generic string backgrounds. If one insists of
using a modification of the patching condition as in (\ref{3tras}), then one has to prove that the string background admits an atlas
such that at all triple overlaps $da^0$ has constant components. It is not apparent that such an atlas exists for general manifolds with a closed 3-form. So the modification
of the patching conditions (\ref{3tras})
may be limited to spaces with toroidal topology.

To summarize, the construction of the double space for  $T^3$ with $H$ flux background depends on  the particular atlas
we have chosen on $T^3$. Therefore, the whole construction is rather special  attached to the details of a manifold structure on  $T^3$. Since it depends
on the choice of the atlas, the construction  is not general  covariant and as a result (\ref{3tras}) does not generalize to all
 spaces in a way similar to (\ref{tras21}).

\section{Patching k-forms and exceptional coordinates}

\subsection{Transition functions of closed k-forms}

The patching conditions of any closed k-form $\omega^k$, $k>3$, on a manifold $M$ can be found in a similar way as those for closed 3-forms in section \ref{patch32}.
For this, it is convenient to use the difference operator $\delta$ of the $\check{\mathrm C}$ech-de Rham theory defined as
\bea
\delta \chi_{\alpha_0\alpha_1\dots \alpha_{p+1}}=\sum_{i=0}^{p+1} (-1)^i \chi_{\alpha_0\alpha_1\dots \hat \alpha_i\dots \alpha_{p+1}}~,
\eea
where $\chi$ is a q-form defined at $p+1$-overlaps and the caret denotes omission. $\delta \chi$ is a q-form defined at (p+2)-overlaps and it is
understood that in the right-hand-side of the above equation $\chi$ is restricted on $U_{\alpha_0}\cap\dots\cap U_{\alpha_{p+1}}$.
Observe that $\delta^2=0$ and $d\delta=\delta d$. For more details on the properties of $\delta$ see eg \cite{bott}.

Given now a globally defined closed k-form on $M$, $k\geq 3$, we use the Poincar\'e lemma to write $\omega_\alpha^k= dC^{k-1}_\alpha$. Then  we obtain the patching conditions\footnote{
There are some sign differences in the definitions of $a$'s as compared with those given  in  section 3.}
\bea
\delta C^{k-1}_{\alpha_0\alpha_1}= d a^{k-2}_{\alpha_0\alpha_1}~,~~~\delta a^{k-2}_{\alpha_0\alpha_1\alpha_2}= d a^{k-3}_{\alpha_0\alpha_1\alpha_2}~,~~~
\label{patchk1}
\eea
and so on till
\bea
\delta a^1_{\alpha_0 \dots \alpha_{k-1}}= d a^0_{\alpha_0 \dots \alpha_{k-1}}~,~~~\delta a^0_{\alpha_0 \dots \alpha_{k}}= n_{\alpha_0 \dots \alpha_{k}}~,
\eea
where $n_{\alpha_0 \dots \alpha_{k}}$ are constants. If $n_{\alpha_0 \dots \alpha_{k}}\in 2\pi\bZ$, then $\omega^k$ represents a class in $H^k(M, \bZ)$.

The (k-1)-form potentials $\{C^{k-1}_\alpha\}$ are not uniquely defined. In particular, there are defined up to a gauge transformation as
\bea
C^{k-1}_\alpha\rightarrow C^{k-1}_\alpha+ d \chi_\alpha^{k-2}~.
\eea
Similarly, the remaining patching conditions are defined up to  gauge transformations of the type
\bea
a^{p}_{\alpha_0\dots \alpha_q}\rightarrow a^{p}_{\alpha_0\dots \alpha_q}+  d\chi^{p-1}_{\alpha_0\dots \alpha_q}+ (\delta \psi^p)_{\alpha_0\dots \alpha_{q}}
\label{trak}
\eea
for some $\chi^{p-1}_{\alpha_0\dots \alpha_q}$ and $\psi^p_{\alpha_0\dots \alpha_{q-1}}$, where $p+q=k-1$. This ambiguity in the definition of patching conditions
of $\omega^k$
is the only one allowed consistent with  $d\omega^k=0$ and the transition functions of $M$.

\subsection{Exact k-forms and patching conditions}

As in the 3-form case utilizing the ambiguity in the definition of the patching conditions (\ref{trak}), it is possible to show  that if  $a^p$, $p=0, \dots, k-2$ satisfies the cocycle
condition
\bea
(\delta a^{p})_{\alpha_0\dots \alpha_{q+1}}=0~,
\eea
then $\omega^k$ is exact.

In what follows, it suffices to prove this for the $a^{k-2}_{\alpha_0\alpha_1}$ as those are responsible for the patching a k-form at double overlaps,
and potentially can be used to construct the exceptional generalized spaces. The proof is similar to that we have given for 3-forms. In particular, suppose that
\bea
(\delta a^{k-2})_{\alpha_0\alpha_1\alpha_2}=0~.
\label{cocycle2}
\eea
Then define
\bea
\tilde C^{k-1}_{\alpha_0}= C^{k-1}_{\alpha_0}- d\big(\sum_{\gamma} \rho_\gamma a^{k-2}_{\gamma \alpha_0}\big)~.
\eea
Clearly $d \tilde C^{k-1}_{\alpha_0}= dC^{k-1}_{\alpha_0}= \omega_{\alpha_0}^k$ as $\tilde C^{k-1}$ and $ C^{k-1}$ are related
up to a gauge transformation.

Moreover $\tilde C^{k-1}$ is a globally defined (k-1)-form as
\bea
\delta( \tilde C^{k-1})_{\alpha_0\alpha_1}&=& (\delta C^{k-1})_{\alpha_0\alpha_1}-d\big(\sum_{\gamma} \rho_\gamma (a^{k-2}_{\gamma \alpha_0}-a^{k-2}_{\gamma \alpha_1})\big)
\cr
&=& d a^{k-2}_{\alpha_0\alpha_1}- d \sum_{\gamma} (\rho_\gamma a^{k-2}_{\alpha_0\alpha_1})=0~,
\eea
where we have used (\ref{cocycle2}) and that $\sum_\gamma \rho_\gamma=1$.

\subsection{Seeking a consistent patching for EFT}

For EFT there is not an analogue of the patching conditions of \cite{hz} available for DFT. Instead the constructions have been based on using
infinitesimal symmetries generated by generalized Lie derivatives, see eg \cite{perry3, waldram2,  berman,perry1, cederwall, samtleben, grana} for  detailed descriptions. One expects
that whatever the final form of the finite transformations are for EFT, these will generate both
the transition functions of the underlying spacetime and the patching conditions of the form field strengths  of the theory. After solving the
strong section condition, the two must be related. Following the analogous  analysis for DFT,  one may hypothesize that the transition functions of the exceptional space read as
\bea
x^i_\alpha=x_{\alpha\beta}(x^j_\beta)~,~~~y^{k-2}_{\alpha_0}-y^{k-2}_{\alpha_1}=- \zeta^{k-2}_{\alpha_0\alpha_1}~,
\label{patchk}
\eea
where $x$ and  $y^{k-2}$  are the spacetime and additional coordinates, respectively, and the
patching data $a^{k-2}_{\alpha\beta}$ of the k-form field strength are a linear combination of $\zeta^{k-2}_{\alpha_0\alpha_1}$. If this is the case,  then again consistency at triple overlaps will require that $\omega^k$ is exact.

In appendix A, we propose a modification of the patching conditions (\ref{patchk}) which resolves the restriction at triple
overlaps.  However, it does not topologically geometrize the k-form field strength. Similar constructions can be made in theories
that we include the dual fields as demonstrated in appendix A for 11-dimensional supergravity.
In section 6, we make an alternative proposal how a priori topologically geometrize k-forms
field strengths based on K-theory and homotopy theory.

The above result does not hold for exceptional generalized geometries, ie those that no new coordinates are introduce in addition to those of spacetime. This is because they do not require the
condition (\ref{cocycle2}) but instead
\bea
d\delta (a^{k-2})_{\alpha_0\alpha_1\alpha_2}=0~.
\eea
This  does not introduce a restriction on the patching conditions of form field strengths.

\section{Summary and outlook}

We have shown that the patching conditions of DFT as arise from generalize coordinate transformations after solving the strong section condition  imply that the NS-NS 3-form field is an exact 3-form. A similar conclusion may hold in the context of EFT under some plausive assumptions regarding the relation between the transition functions of additional exceptional coordinates
and the patching data of the form field strengths.
We have also explored some alternative possibilities. These resolve some of the difficulties, like the restriction on the form field strengths to be exact,
but they do not obey the topological geometrisation condition which is one  of the key properties that the $U(1)$ field paradigm.

Furthermore we revisited the double space construction of $T^3$ with $H$-flux background from the patching conditions point of view.
If the combinatorial law of the transition functions is not altered at triple overlaps, the construction is inconsistent. However, there
is a modification of the combinatorial law at triple overlaps which allows for a consistent construction of the double space. But
this modification depends on the choice of an atlas on $T^3$, ie it is not general covariant, and so as a result cannot be generalized
to generic backgrounds with $H$-fluxes.

To introduce new coordinates that extend the spacetime in a consistent way without any further conditions on the fields, like exactness of the form field strengths, one may
try to generalize some aspects of  the 2-form paradigm reviewed in section 2. Some of the directions that can be pursued are the following.

\begin{itemize}

\item To modify the combinatorial law of transition functions.

\item To modify the transition functions.

\item To introduce topology on the generalized spaces.

\end{itemize}

It is clear that the 2-form paradigm is consistent because it has been possible to appropriately modify the combinatorial law of the transition
functions. Of course, this has been achieved under the additional requirement that $\omega^2$ represents a class in
$H^2(M, \bZ)$.  Although this  imposes a restriction on the transition functions of $\omega^2$, this restriction is required by the Dirac quantisation condition.
 In the context of DFTs and EFTs, this is an indication that non-commutative geometry has a role   as it is not possible to alter the combinatorial law
of transition functions in a straight forward way, see also \cite{bakas}.

As we has seen a mild modification of the patching conditions of the form field strengths, eg as linear functions of the transition functions of the additional coordinates,
leads to the conclusion that consistency at triple overlaps requires that the form field strengths are exact.
However as we explain in appendix A, there is a modification of the transition functions of the additional coordinates such that the double and exceptional
spaces are consistent at triple overlaps without any restrictions on the patching conditions of the form field strengths.
However, as such a modification has its problems,  like for example the topological geometrisation condition does not hold. Moreover there is no use of the cohomological analogue
the Dirac quantisation condition  $[\omega^k]\in H^k(M,\bZ)$ in the construction of the double of exceptional space above which has a central role
in the $U(1)$ field paradigm.

It is clear from the above that whatever the construction of these extended spaces is the additional coordinates have to have a non-trivial
topology. This is the only way that both the topological geometrisation  and the cohomological analogue of the Dirac quantisation  $[\omega^k]\in H^k(M,\bZ)$ conditions
can be utilized to construct these spaces. It is not a priori apparent how this can be done or whether a consistent construction is possible for all cases of interest,
beyond those of toroidal compactifications, but a way to proceed is as follows.

One of the difficulties in adapting the $U(1)$ field paradigm in the context of string theory and M-theory is that after applying a duality transformation
the spacetime may change as a manifold at the same time as the fields of the theory. One way to incorporate this into the construction of extended spaces is as follows.
Suppose that $(M_I, {\cal F}_I)_{I\in {\cal I}}$ be a family of spacetimes  $M_I$ with field content ${\cal F}_I$ such that any two pairs $(M_I, {\cal F}_I)$
and $(M_J, {\cal F}_J)$ are related by a duality transformation $D_{IJ}$, $D_{IJ}:~ (M_I, {\cal F}_I)\rightarrow (M_J, {\cal F}_J)$.
One way to  geometrize  the data  $(M_I, {\cal F}_I)$ and  $D_{IJ}$ is to assume that there is a space $CM$, a C-fold,  and maps $\pi_I: CM\rightarrow M_I$
such that
\begin{enumerate}

\item  $(\pi_I)_* TCM=TM_I$,

\item  the duality transformation $D_{IJ}$ has a lift $\tilde D_{IJ}$ on $CM$ which is implemented
with a transformation which respects the topological and geometric properties of the C-fold, $CM$,  and $ D_{IJ}\circ \pi_I= \pi_J\circ \tilde D_{IJ}$, and

\item  as a minimal requirement
assert that the pull back $\pi_I^* \omega_I$ of all the fields $\omega_I\in {\cal F}_I$ on $M_I$ which represent non-trivial classes $[\omega_I]$ in cohomology $H^*(M_I)$
represent  the trivial class $[ \pi_I^* \omega_I]=0$ in $H^*(CM)$.

\end{enumerate}
The last property
is the implementation of the topological geometrisation condition which should appear as a weak restriction  for carrying out the
geometrisation programme. It would be necessary to impose additional conditions on $CM$  but the above three conditions can serve
as a minimal requirement.

It is not apparent whether the construction of  C-folds $CM$ would be possible for all backgrounds  in string theory and M-theory. However, there are constructions which satisfy some of the three conditions mentioned above.  For this consider the trivial
case, where the duality transformation $D_{IJ}$ is the identity and let us view $CM$ as a fibration over $M$. There are many ways
that  the construction of a $CM$ can be achieved which satisfies the topological geometrisation property. One way is K-theory which generalizes the 2-form paradigm.  Suppose that one tries to construct a C-fold, $CM_{\omega^4}$, of a spacetime that has
 field content a closed 4-form $\omega^4$. If $\omega^4$ is represented by the second Chern class or the first Pontryagin class of a complex or real vector bundle $E$, respectively, then
as a $CM$ can be taken as the principal bundle $P$ that $E$ is associated to. In such a case $\omega^4$ will represent an integral class and
the pull-back of $\omega^4$ on $P$ will represent the trivial class. It may be possible to find amongst the K-theory class  which resolves the topological geometrisation condition
a representative which will also exhibit the local geometric requirements as expected from DFTs and EFTs. K-theory has appeared before in string theory and M-theory \cite{witten}, see also \cite{witten2} and references within. Therefore, it may not be a surprise that it could be also applicable in this context.
Moreover the tangent space of all fibrations are  extensions of that of the base space,  and so $T(CM_{\omega^4})\rightarrow \pi^*TM\rightarrow 0$. This is
one of the properties expected for generalized manifolds.

It is tantalizing for example that one can construct C-folds for the  $AdS_7\times S^4$ near horizon geometry of M5-branes.
 It is known that for every class in $[\omega]\in H^4(S^4, \bZ)$ there is a complex vector bundle $E$ such that $c_2(E)=[\omega]$.
 This follows from the isomorphism of $\tilde K(S^4)$ with $H^4(S^4, \bZ)$, see eg \cite{huse}.  The associated real bundle is an $SO(4)$ bundle
 with Pontryangin class $p_1(E_{\bR})=2 c_2(E)$. In fact all the $SO(4)$ bundles are classified by two integers $n,m$ as
$\pi_3(SO(4))=\bZ\oplus \bZ$ and $p_1=2 (n-m)$ \cite{milnor}.  Considering the associated $SO(4)$ principal bundles,
observe  that $T_eSO(4)=\mathfrak{so}(4)=\Lambda^2(\bR^4)$.  So the tangent space of the fibres
provide a  local model for the  2-forms on  $S^4$. This is in line with the expectation that the additional coordinates of this example
are locally modeled by 2-forms on  $S^4$.

Furthermore, there are homotopy constructions which can implement the topological geometrisation property.  One such construction
is the Whitehead towers, see eg \cite{bott}. These are given by successive fibrations
\bea
K(\pi_n, n-1)\rightarrow Y_{n+1}\rightarrow Y_n~,
\eea
where $\pi_n=\pi_n(X)$,  $Y_1=X$, $\pi_k(Y_n)=0$ for $k<n$ and $\pi_k(Y_n)=\pi_k(X)$ for $k>n$, and $K(\pi_n, n-1)$ are Eilenberg-MacLane spaces.  Therefore, $Y_2$ is the universal cover of $X$,
$\pi_1(Y_3)=\pi_2(Y_3)=0$ and so on.  Thus if $[\omega]$ represents a class in the first non-vanishing cohomology group of $X$, say $H^n(X, \bZ)$,
this class becomes trivial in $Y_{n+1}$ implementing the topological geometrisation condition.

Although the homotopy constructions above based on the Whitehead towers can implement the topological geometrisation
condition, the spaces involved are  infinite dimensional.   It is not apparent that one can generically
 find a finite dimensional model which has all the required properties as for the $AdS_7\times S^4$
 example presented above. Nevertheless the above homotopy constructions use path spaces and so may admit a
 string theoretic interpretation.

This procedure based on Whitehead towers can be applied to implement the topological geometrisation condition
 on  a Calabi-Yau  mirror pair.  Suppose $X_1$ is a Calabi-Yau
manifold and $X_2$ its mirror.  Consider $X_1\times X_2$. The first non-vanishing cohomology group is
$H^2(X_1\times X_2)=H^2(X_1, \bZ)\oplus H^2(X_2,\bZ)$. To implement the topological geometrisation condition, it suffices
to consider $Y_4$.  By construction $H^2(Y_4, \bZ)= H^3(Y_4, \bZ)=0$ and $Y_4$ fibres above $X_1\times X_2$ and so above both $X_1$ and $X_2$.
Therefore the $H^2$ and $H^3$ cohomologies classes of both $X_1$ and $X_2$ become trivial in $Y_4$. It will be of interest to
determine the action of $\tilde D_{12}$ on $Y_4$, where $D_{12}$ is the mirror symmetry of
 $X_1$ and $X_2$.

Therefore there are some suggestions how to construct generalized manifolds which allow for form field strengths to represent non-trivial cohomology classes. However,
it is less apparent which will be the most fruitful way to proceed. The expectation is that the full theory at the end will combine many of the local computations
that have been done so far with the global aspects that many of the backgrounds have in a consistent way.

\vskip 1cm

\noindent{\bf Acknowledgements} \vskip 0.1cm
I would like to thank  Martin Cederwall,  Ulf Gran and specially  David Berman for many helpful comments. I am partially supported by the STFC grant ST/J002798/1.

\vskip 0.5cm
{\small \noindent{\bf Note added:}
Part of the title of this paper was inspired from an article by  Greg Child contained in his book ``Mixed Emotions'', published by {\it The Mountaineers}, Seattle Washington (1993).}

\setcounter{section}{0}\setcounter{equation}{0}

\appendix

\section{Modifying transition functions}
\subsection{New transition functions and k-forms}

Let $\omega^k$ be  closed k-form on $M$, $k\geq 3$, with transition functions as those defined in section 4.
To define $\tilde M_{\omega^k}$, we introduce  coordinates $(x, y^{k-2})$, where $x$ are the coordinates of $M$, and impose the transition functions
\bea
x_{\alpha_0}=f_{\alpha_0\alpha_1}(x_{\alpha_1})~,~~~\delta y^{k-2}_{\alpha_0\alpha_1}\equiv- y^{k-2}_{\alpha_0}+y^{k-2}_{\alpha_1}= a^{k-2}_{\alpha_0\alpha_1}-\sum_{\beta} \rho_\beta\,\, d a^{k-3}_{\beta\alpha_0\alpha_1}~.
\label{patchkk}
\eea
Since $\delta^2=0$, consistency at triple overlaps requires that
\bea
\delta\big ( a^{k-2}-\sum_{\beta} \rho_\beta\,\, d a^{k-3}_{\beta} \big)_{\alpha_0\alpha_1\alpha_2}=0
\eea
which is satisfied as it can be shown after a calculation using (\ref{patchk1}).

It is clear that there is a projection $\pi$ from $\tilde M_{\omega^k}$ onto $M$ and so it can be thought that $\tilde M_{\omega^k}$ is a bundle over $M$.
However as in the case of 3-forms there is no natural global section of $M$ in $\tilde M_{\omega^k}$, eg the zero section is not preserved by the transition functions.

Since $\tilde M_{\omega^k}$ is a manifold, one can investigate the tangent as well as all the other tensor bundles in the standard way.
We shall do this for $\tilde M_{\omega^3}$ as the generalization to $\tilde M_{\omega^k}$ is straightforward.  To do this
let us write the patching conditions (\ref{patchk}) explicitly as
\bea
x^i_\alpha=f^i_{\alpha\beta}(x^j_\beta)~,~~~(y^1_\alpha)_{i}- (y^1_\beta)_{k}
{\partial x_\beta^k \over \partial x^i_\alpha}= (\tilde a^1_{\alpha\beta})_{i}
\label{2patch}
\eea
where
\bea
\tilde a^1_{\alpha\beta}=-a^1_{\alpha\beta}+\sum_{\delta} \rho_\delta\,\, d a^0_{\delta\alpha\beta}~.
\eea
Note that the second patching condition in (\ref{2patch}) does not satisfy the strong section condition. However one can do a coordinate redefinition
and set simply $(y^1_\alpha)_{i}=(y^1_\beta)_{i}+(\tilde a^1_{\alpha\beta})_{i}$ but this is less attractive in calculations.

To find the  patching conditions of the tangent bundle consider the vector field $X_\alpha= A_\alpha^i {\partial\over \partial x^i_\alpha}+ B_{\alpha, i} {\partial\over \partial (y_\alpha)_{i}}$
and demanding $X_\alpha=X_\beta$, we find that
\bea
A_\alpha^i&=&{\partial x^i_\alpha\over \partial x^j_\beta} A^j_\beta~,~~~
\cr
(B_\alpha)_{i}&=&{\partial x_\beta^k \over \partial x^i_\alpha}
 (B_\beta)_{k}+  {\partial x_\alpha^m \over \partial x^p_\beta}{\partial^2 x_\beta^k \over \partial x^m_\alpha \partial x^i_\alpha}
 (y_\beta)_{k} A^p_\beta +{\partial (\tilde a^1_{\alpha\beta})_{i} \over \partial x_\beta^k} A^k_\beta~.
\eea
It is clear from this that $T \tilde M_{\omega^3}$ is an extension of $\pi^* TM$ with respect to the cotangent bundle $\pi^* \Lambda^1M$ of $M$, ie
\bea
0\rightarrow \pi^* \Lambda^1M \rightarrow T \tilde M_{\omega^3}\rightarrow \pi^* TM\rightarrow 0~.
\eea
In particular, $\pi^* \Lambda^1M$ is a subbundle of $T\tilde M_{\omega^3}$.  This is reminiscent of generalized geometry where
the bundle $E$ over $M$ which is an extension of $TM$ is now replaced with  $T \tilde M_{\omega^3}$.

Furthermore observe that the  pairing of $TM$ and $\Lambda^1M$ is naturally extended to $T \tilde M_{\omega^3}$ as the transition
functions of $T \tilde M_{\omega^3}$ preserve it. As a result, one can define an $O(n,n)$ structure on $T \tilde M_{\omega^3}$ as expected from string theory considerations
which  also arises in the context of generalized geometry. But of course this $O(n,n)$ is related to the modified transition functions rather than those
associated to the original patching conditions $a^1_{\alpha\beta}$ of of the 2-form gauge potential.

In the more general case of $\tilde M_{\omega^k}$, one finds that
\bea
0\rightarrow \pi^*\Lambda^{k-2} M \rightarrow TM_{\omega^k} \rightarrow \pi^* TM\rightarrow 0~.
\eea

\subsection{Testing for other properties}

One of the properties of the construction of C-folds for 2-forms $\omega^2$ is that when one pulls back the 2-form on the C-fold, $\omega^2$ becomes exact. It is not expected that this property
holds on $\tilde M_{\omega^k}$ because by construction the fibres have trivial topology, ie by construction $\tilde M_{\omega^k}$ is contractible to $M$. Nevertheless,
it is instructive to see what the result is. In analogy with the 2-form case, we take the exterior derivative of second transition function in (\ref{patchk}) to find
\bea
dy^{k-2}_\alpha-dy^{k-2}_\beta= C^{k-1}_\alpha-C^{k-1}_\beta+ \sum_{\gamma} d\rho_\gamma \wedge d a^{k-3}_{\alpha\beta\gamma}~.
\eea
Using the properties of the partition of unity and the patching conditions of $\omega^k$, this can be rewritten as
\bea
dy^{k-2}_\alpha-C^{k-1}_\alpha-\sum_{\gamma} d\rho_\gamma\wedge a^{k-2}_{\gamma\alpha}=dy^{k-2}_\beta-C^{k-1}_\beta-\sum_{\gamma} d\rho_\gamma\wedge a^{k-2}_{\gamma\beta}~.
\eea
It is clear from the above equation that $\omega^k$ cannot be written as the exterior derivative of a $(k-1)$-form on $\tilde M_{\omega^k}$.  For the latter,
one would have expected that $dy^{k-2}_\alpha-C^{k-1}_\alpha$ should have patched globally as a (k-1)-form on $\tilde M_{\omega^k}$. But as we have shown
this is not the case and the (k-1)-form that patches globally receives a correction that depends on the derivative of the functions $\rho_\alpha$ which appear
in the partition of unity.    This correction is like a source term. Assuming that the partition of unity functions have compact support and that the good cover is very fine,
$\rho_\alpha$ resemble delta-functions.  Therefore the source term is like the derivative of a delta function. Note that in the construction of some exceptional field theories additional field are required in
the form of tensor  hierarchies, see \cite{samtleben, grana} which is reminiscent to these additional terms.

\subsection{$\tilde M$ for M-theory}

It would be of interest in view of applications in strings and M-theory to generalize the construction of the previous appendix from single k-forms, $\omega^k$, to differential
algebras ${\cal A}$. We shall not give a general treatment of this. Instead we shall focus on the differential algebra of M-theory generated by the 4-form field strength $F$ and its dual $G$, where
now
\bea
{\cal A}:~~~dF=0~,~~~dG= F\wedge F~,
\label{malg}
\eea
and $G$ is treated as an independent field.

Suppose that $M$ now has a good cover, the above equations can be solved at each open set of the cover as
\bea
F_\alpha= dC^3_\alpha~,~~~ G_\alpha= dC^6_\alpha+ C_\alpha^3\wedge F_\alpha
\eea
Next on double overlaps, we have
\bea
\delta C^3_{\alpha\beta}= da^2_{\alpha\beta}~,~~~\delta C^6_{\alpha\beta}= d b^5_{\alpha\beta}- a^2_{\alpha\beta}\wedge F
\eea
and at triple overlaps
\bea
\delta a^2_{\alpha\beta\gamma}=d a^1_{\alpha\beta\gamma}~,~~~\delta b^5_{\alpha\beta\gamma}=d b^4_{\alpha\beta\gamma}+ a^1_{\alpha\beta\gamma}\wedge F
\eea
Next to construct $\tilde M$, we introduce coordinates $(x, y^2, w^5)$ and introduce the transition functions
\bea
x_{\alpha}&=&f_{\alpha\beta}(x_{\beta})~,~~~y^{2}_{\alpha}-y^{2}_{\beta}= a^{2}_{\alpha\beta}-\sum_{\gamma} \rho_\gamma\,\, d a^1_{\gamma\alpha\beta}~,
\cr
w^5_{\alpha}-w^5_\beta&=& b^5_{\alpha\beta}-\sum_{\gamma} \rho_\gamma\,\, (a^1_{\gamma\alpha\beta}\wedge F+ d b^4_{\gamma\alpha\beta})~.
\label{patchm}
\eea
After performing a computation similar to that we have explained in previous case, the transition functions are consistent at triple overlaps.
This proves that one can define a manifold $\tilde M_{\cal A}$ for the M-theory differential algebra ${\cal A}$ in (\ref{malg}).

Most of the properties of the $\tilde M_{\omega^k}$ spaces constructed for single k-forms can be generalized to this case. First there is a projection from $\tilde M_{\cal A}$
onto $M$, and so $\tilde M_{\cal A}$ can be thought as a bundle over $M$. The tangent space $T\tilde M_{\cal A}$ is an extension of $TM$ with respect to the space of
2- and 5-forms on $M$, ie
\bea
0\rightarrow \pi^*\Lambda^2 M\oplus \pi^*\Lambda^5 M\rightarrow T\tilde M_{\cal A}\rightarrow \pi^* TM\rightarrow 0
\eea
As this construction is intended as an application to M-theory, $M$ is 11-dimensional. However a similar construction can be applied to compactifications.

\setcounter{section}{0}\setcounter{equation}{0}

\end{document}